\documentclass[
    ,final            % use final for the camera ready runs
  ]
  {aipproc}

\layoutstyle{8x11single}
\usepackage{graphicx,color}

\begin{document}

\title{Renormalization Group Invariance in the Subtractive Renormalization Approach to the $NN$ Interactions}


\author{S. Szpigel}{
  address={Faculdade de Computa\c c\~ao e Inform\'atica, Universidade Presbiteriana Mackenzie, S\~ao Paulo, SP, Brasil}
}

\author{V. S. Tim\'oteo}{
  address={Grupo de \'Optica e Modelagem Num\'erica-GOMNI, Faculdade de Tecnologia, Universidade Estadual de Campinas-UNICAMP, Limeira, SP, Brasil}
}

\begin{abstract}
In this work, we apply the subtracted kernel method (SKM) to the chiral nucleon-nucleon ($NN$) interaction in the $^3P_0$ channel up to next-to-next-to-leading-order ($NNLO$). We demonstrate, by explicit numerical calculations, that the SKM procedure is renormalization group invariant under the change of the subtraction scale provided the driving-term of the subtracted scattering equation is evolved through a non-relativistic Callan-Symanzik (NRCS) flow-equation.
\end{abstract}

\maketitle

%%%%%%%%%%%%%%%%%%%%%%%%%%%%%%%%%%%%%%%%%%%%
%% MAINMATTER
%%%%%%%%%%%%%%%%%%%%%%%%%%%%%%%%%%%%%%%%%%%%

\section{Introduction}

The standard approach to the non-perturbative renormalization of nucleon-nucleon ($NN$) interactions in the context of chiral effective field theory (ChEFT) can be divided in two steps \cite{epelbaum1a}. In the first step, one has to solve a regularized Lippmann-Schwinger (LS) equation for the scattering amplitude by iterating the effective $NN$ potential truncated at a given order in the chiral expansion, which includes long-range contributions from pion-exchange interactions and short-range contributions parametrized by nucleon contact interactions. The most common scheme used to regularize the ultraviolet (UV) divergences in the LS equation is to introduce a sharp or smooth momentum cutoff that suppresses the contributions from the potential matrix-elements for momenta larger than a given momentum cutoff scale (multi-pion exchange interactions also involve UV divergent loop integrals which must be consistently regularized and renormalized) \cite{epelbaum1a,machleidt1}. In the second step, one has to determine the strengths of the contact interactions, the so called low-energy constants (LEC's), by fitting a set of low-energy scattering data. Once the LEC's are fixed at a given momentum cutoff scale, the effective $NN$ potential can be used to evaluate other observables. The $NN$  interactions can be considered properly renormalized when the predicted observables are (approximately) renormalization group invariant (i.e., independent of the momentum cutoff scale) within the range of validity of ChEFT \cite{machleidt1,lepage}.

An alternative approach to the non-perturbative renormalization of $NN$ interactions is the subtracted kernel method (SKM) \cite{npa99,plb00,hep01,plb05,npa07,ijmpe07,prc11,fb19,efb21,aop2010,jpg2012}, which is based on recursive multiple subtractions performed in the kernel of the scattering equation at a given energy scale.  In this work, we apply the SKM to the chiral $NN$ interaction in the $^3P_0$ channel up to next-to-next-to-leading-order ($NNLO$) and demonstrate, by explicit numerical calculations, that the SKM procedure is renormalization group invariant under the change of the subtraction scale provided the driving-term of the subtracted scattering equation is evolved through a non-relativistic Callan-Symanzik (NRCS) flow-equation.

\section{SKM approach for the NN system}

We start by considering the chiral expansion for the effective $NN$ potential in Weinberg's power counting scheme (WPC) \cite{epelbaum1a,machleidt1}. In a partial-wave relative momentum space basis, the matrix-elements of the $NN$ potential in the $^3P_0$ channel up to $NNLO$ are given by
\begin{eqnarray}
V^{LO}(p,p') &=& V^{LO}_{1\pi}(p,p') \;,\nonumber\\
V^{NLO}(p,p') &=& V^{LO}(p,p') + V^{NLO}_{1\pi}(p,p')+ V^{NLO}_{2\pi}(p,p')+C_1~p~p' \;,\nonumber\\
V^{NNLO}(p,p') &=& V^{NLO}(p,p')+ V^{NNLO}_{1\pi}(p,p')+ V^{NNLO}_{2\pi}(p,p')\; ,
\end{eqnarray}
\noindent
where the coefficient $C_1$ stands for the strength of the $NLO$ contact interaction in the $^3P_0$ channel and an obvious notation is used for the pion-exchange interactions.

The leading-order ($LO$) unprojected one-pion exchange potential (OPE) is given by
\begin{eqnarray}
V^{LO}_{1\pi}(\vec p, \vec {p'})=-\frac{g_a^2}{4(2\pi)^3f_\pi^2} \vec\tau_1\cdot\vec\tau_2
\frac{\vec\sigma_1\cdot(\vec {p'}-\vec {p})\;
\vec\sigma_2\cdot(\vec {p'}-\vec {p})}
{ (\vec{p'}-\vec{p})^2+m_\pi^2} \; ,
\end{eqnarray}
\noindent
where $\vec\tau_i$ and $\vec\sigma_i$ are the isospin and spin Pauli operators, $g_a$, $f_\pi$ and $m_\pi$ denote, respectively, the axial coupling constant, the pion weak-decay constant and the pion mass. The higher-order OPE terms include corrections from pion loops and counter term insertions, which only contribute to the renormalization of coupling constants and masses. In this work, we use $g_a=1.25$, $f_\pi =93~{\rm MeV}$ and $m_\pi=138~{\rm MeV}$. The two-pion-exchange (TPE) potential is taken from Ref. \cite{epelbaum2}.

%%%%%%%%%%
Consider the LS equation for the $T$-matrix of a two-nucleon system, which can be written in operator form as
\begin{eqnarray}
T(E) &=& V + V~G_{0}^{+}(E)~T(E) \; ,
\label{LS}
\end{eqnarray}
where $E$ is the energy of the two-nucleon system in the center-of-mass frame, $V$ is the effective $NN$ potential and $G_{0}^{+}(E)= (E - H_{0} + i \epsilon)^{-1}$ is the free Green's function for the two-nucleon system with outgoing-wave boundary conditions, given in terms of the free hamiltonian $H_0$. Both pion-exchange and contact interaction terms can lead to UV divergences when the effective $NN$ potential $V$ at a given order in the chiral expansion is iterated in the LS equation, requiring a regularization and renormalization procedure in order to obtain well-defined finite solutions.
%%%%%%%%%%%

In the standard cutoff renormalization scheme the formal LS equation, Eq. (\ref{LS}), is regularized by multiplying the effective $NN$ potential $V$ with a momentum cutoff regularizing function. The common choice is an exponential $f(p)=\exp[-(p/\Lambda)^{2r})]$ (with $r=1, 2, \ldots$), where $\Lambda$ is a cutoff parameter, such that
\begin{equation}
 V(p, p') \rightarrow  V_{\Lambda}(p, p') \equiv \exp[-(p/\Lambda)^{2r})]~ V(p, p')~ \exp[-(p'/\Lambda)^{2r})] \; .
 \label{CR}
\end{equation}

In the SKM approach, a regularized and renormalized LS equation for the $T$-matrix at a given order in the chiral expansion is computed through an iterative procedure which involves recursive multiple subtractions in kernel. For a general number of subtractions $n$, we define a $n$-fold subtracted LS equation given in operator form by
\begin{equation}
T(E) = V^{(n)}_{\mu}(E) + V^{(n)}_{\mu}(E)~G_{n}^{+}(E;-\mu^2)~T(E) \; ,
\label{LSn}
\end{equation}
where $\mu$ is the subtraction scale and $E$ is the energy of the two-nucleon system in the center-of-mass frame. The $n$-fold subtracted Green's function is defined by
\begin{equation}
G_{n}^{+}(E;-\mu^2) \equiv \left[(-\mu^2-E)~ G_{0}^{+}(-\mu^2) \right]^{n}~G_{0}^{+}(E)  \; ,
\label{Gn}
\end{equation}
\noindent
where $G_{0}^{+}(E)$ is the free Green's function. Note that we choose a negative energy subtraction point $-\mu^2$, such that the free Green's function $G_{0}^{+}(-\mu^2)$ is real. For convenience, here we implement the SKM procedure using the $K$-matrix instead of the $T$-matrix. The LS equation for the partial-wave $K$-matrix with $n$ subtractions is given by
\begin{eqnarray}
K^{(n)}_{\mu}(p,p';k^2) = V^{(n)}_{\mu}(p,p';k^2) + \frac 2 \pi \mathcal{P} \int_0^\infty dq~q^2
\left(\frac{\mu^2+k^2}{\mu^2+q^2}\right)^n \frac{V^{(n)}_{\mu}(p,q;k^2)}{k^2-q^2}K^{(n)}_{\mu}(q,p';k^2) \; ,
\label{subkn}
\end{eqnarray}
\noindent
where $k=\sqrt{E}$ is the on-shell momentum in the center-of-mass frame and ${\cal P}$ denotes the principal value. Note that the $n$-fold subtracted Green's function introduces an energy- and $\mu$-dependent form factor in the kernel of the subtracted LS equation that regularizes ultraviolet (UV) power divergences up to order $q^{2n-1}$, effectively acting like a smooth momentum cutoff regularizing function. The driving term $V^{(n)}_{\mu}$ at each order is computed through an iterative procedure, starting from the {\it ansatz} for the leading-order ($LO$) driving term given by:
\begin{equation}
V^{(1)}_{\mu}(p,p';k^2) = V^{LO}_{1\pi}(p,p') \; .
\label{v1}
\end{equation}

To obtain the next-to-leading-order ($NLO$) driving term, $V^{(3)}_{\mu}$, we first calculate ${\bar V}^{(2)}$ from $V^{(1)}$,
\begin{eqnarray}
{\bar V}^{(2)}_{\mu}(p,p';k^2) = V^{(1)}_{\mu}(p,p';k^2)- \frac 2 \pi \int_0^\infty dq~q^2 ~ V^{(1)}_{\mu}(p,q;k^2)~\frac{(\mu^2+k^2)^{1}}{(\mu^2+q^2)^2}{\bar V}^{(2)}_{\mu}(q,p';k^2) \; .
\label{v2}
\end{eqnarray}
\noindent
Then, we calculate ${\bar V}^{(3)}$ from ${\bar V}^{(2)}$,
\begin{eqnarray}
{\bar V}^{(3)}_{\mu}(p,p';k^2) = {\bar V}^{(2)}_{\mu}(p,p';k^2)- \frac 2 \pi \int_0^\infty dq~q^2 ~{\bar V}^{(2)}_{\mu}(p,q;k^2)~\frac{(\mu^2+k^2)^{2}}{(\mu^2+q^2)^3}{\bar V}^{(3)}_{\mu}(q,p';k^2) \; ,
\label{v3}
\end{eqnarray}
\noindent
and add the $NLO$ interactions:
\begin{eqnarray}
V^{(3)}_{\mu}(p,p';k^2)= {\bar V}^{(3)}_{\mu}(p,p';k^2) + V^{NLO}_{2\pi}(p,p')+ C_1(\mu)~p~p' \; .
\label{v3nlo}
\end{eqnarray}

To obtain the next-to-next-to-leading-order ($NNLO$) driving term, $V^{(4)}_{\mu}$, we first calculate ${\bar V}^{(4)}$ from $V^{(3)}$,
\begin{eqnarray}
{\bar V}^{(4)}_{\mu}(p,p';k^2) = V^{(3)}_{\mu}(p,p';k^2)- \frac 2 \pi \int_0^\infty dq~q^2 ~ V^{(3)}_{\mu}(p,q;k^2)~\frac{(\mu^2+k^2)^{3}}{(\mu^2+q^2)^{4}}{\bar V}^{(4)}_{\mu}(q,p';k^2) \; ,
\label{v4}
\end{eqnarray}
\noindent
and add the $NNLO$ interactions:
\begin{equation}
V^{(4)}_{\mu}(p,p';k^2)= {\bar V}^{(4)}_{\mu}(p,p';k^2)+V^{NNLO}_{2\pi}(p,p') \; .
\label{v4nnlo}
\end{equation}
\noindent
One should note that in the WPC scheme only a TPE interaction is added at $NNLO$.

The renormalized strengths $C_i(\mu)$ of the contact interactions included in the driving term $V^{(n)}_{\mu}$ at each order in the chiral expansion are fixed at the subtraction scale $\mu$ by fitting data for low-energy scattering observables, thus encoding the input physical information. Instead of the usual matching of scattering data at discrete values of the on-shell momentum $k$, we follow the procedure described by Steele and Furnstahl \cite{steele1,steele2}, which is numerically much more robust. Here, we use as ``data" the values of the inverse on-shell $K$-matrix evaluated from the solution of the LS equation with the Nijmegen-II potential \cite{nijmegen}, $1/K_{{\rm NIJ}}(k,k;k^2)$, for a spread of very small momenta $k~(\le 0.1~{\rm fm^{-1}})$. We evaluate the inverse on-shell $K$-matrix from the solution of the $n$-fold subtracted LS equation, $1/K^{(n)}_{\mu}(k,k;k^2)$, and fit the difference $\Delta(k^{l+l'}/K)=k^{l+l'}/K_{{\rm NIJ}}-k^{l+l'}/K^{(n)}_{\mu}$ to an interpolating polynomial in $k^2/\mu^2$ to highest possible degree:
\begin{eqnarray}
\Delta\left(\frac{k^{l+l'}}{K}\right)=\frac{k^{l+l'}}{K_{{\rm NIJ}}(k,k;k^2)}-\frac{k^{l+l'}}{K^{(n)}_{\mu}(k,k;k^2)}= A_0 + A_2\;\frac{k^2}{\mu^2} + A_4\;\frac{k^4}{\mu^4} + \ldots \;.
\label{deltainvK}
\end{eqnarray}
\noindent
%%%%%%%%%%%%%%%
The coefficients $A_i$ are then minimized with respect to the variations in the renormalized strengths $C_i(\mu)$. In the case of the $NN$ interaction in the $^3P_0$ channel up to $NNLO$, we use Eq.~(\ref{deltainvK}) for $n=4$ and $l=l'=1$, and minimize the coefficient $A_0$ with respect to the variations in the renormalized strength $C_1(\mu)$.

\section{Renormalization group invariance in the SKM approach}

As pointed before, the multiple subtractions performed in the SKM procedure introduce a form factor in the kernel of the LS equation which acts like a regularizing function, such that the subtraction scale $\mu$ ends up playing a role similar to that of a smooth momentum cutoff scale. But the subtraction scale $\mu$ is arbitrary, and so the scattering observables calculated from the solution of the subtracted LS equation for the scattering amplitude should not depend on its particular choice. By requiring the fully off-shell $K$-matrix with $n$ subtractions to be invariant under the change of the subtraction scale $\mu$, a renormalization group (RG) equation can be derived for the driving term $V^{(n)}_{\mu}(E)$ in the form of a non-relativistic Callan-Symanzik (NRCS) flow equation \cite{plb00}, which is given in operator form by
\begin{equation}
\frac{\partial V^{(n)}_{\mu}(E)}{\partial\mu^2} = -V^{(n)}_{\mu}(E)~\frac{\partial G_{n}^{+}(E;-\mu^2)}{\partial\mu^2}~V^{(n)}_{\mu}(E) \; ,
\label{CSEn}
\end{equation}
\noindent
with the boundary condition $V^{(n)}_{\mu}|_{\mu \rightarrow {\bar \mu}}= V^{(n)}_{\bar \mu}$ imposed at some reference subtraction scale $\bar \mu$ where the renormalized strengths of the contact interactions $C_i(\mu)$ are fixed to fit low-energy observables used as physical input.

In order to explicitly demonstrate the renormalization group invariance in the SKM approach, we consider the evolution through the NRCS flow equation of the $^3P_0$ channel driving term at $NNLO$. In a partial-wave relative momentum space basis, the NRCS flow equation for the matrix-elements of the driving term $V^{(n)}_{\mu}$ is given by
\begin{eqnarray}
\frac{\partial V^{(n)}_{\mu}(p,p';k^2)}{\partial\mu^2}=\frac 2 \pi \int_0^\infty dq~q^2\left[n\frac{(\mu^2+k^2)^{n-1}}{(\mu^2+q^2)^{n+1}}\right]
V^{(n)}_{\mu}(p,q;k^2)~V^{(n)}_{\mu}(q,p';k^2) \; .
\label{CSEnPW}
\end{eqnarray}

\begin{figure}[h]$
\begin{array}{cccc}
\includegraphics[scale=0.41]{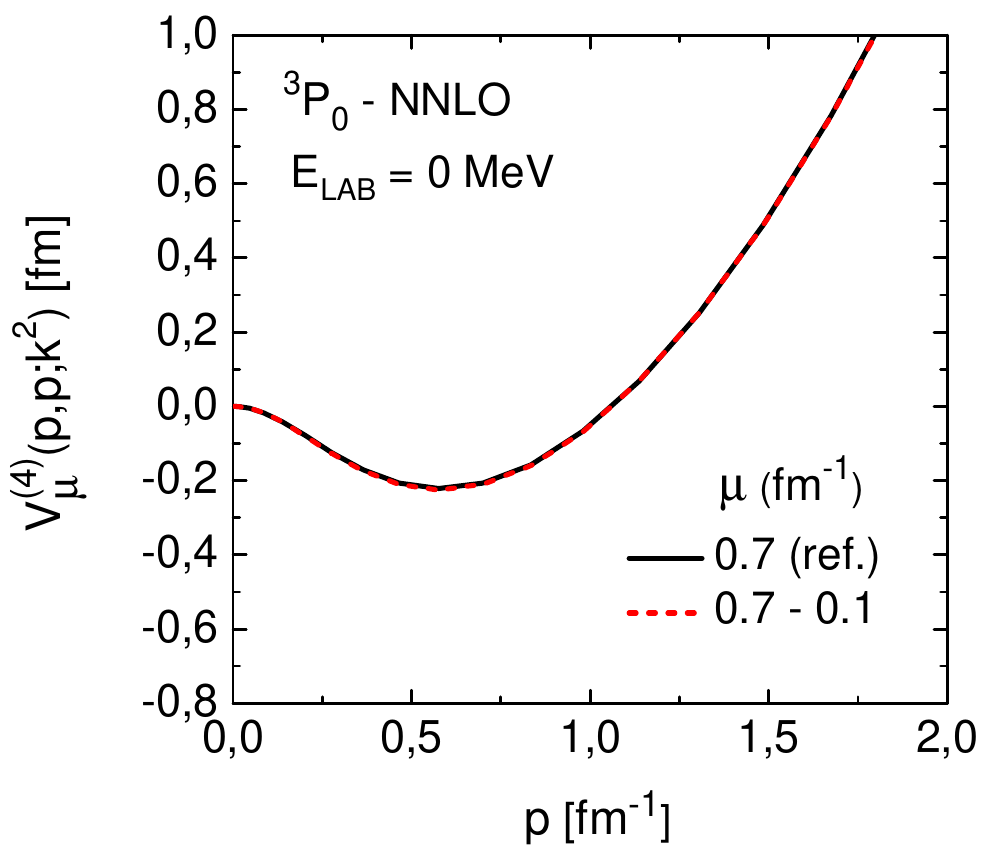}&\hspace{-0.3cm}
\includegraphics[scale=0.41]{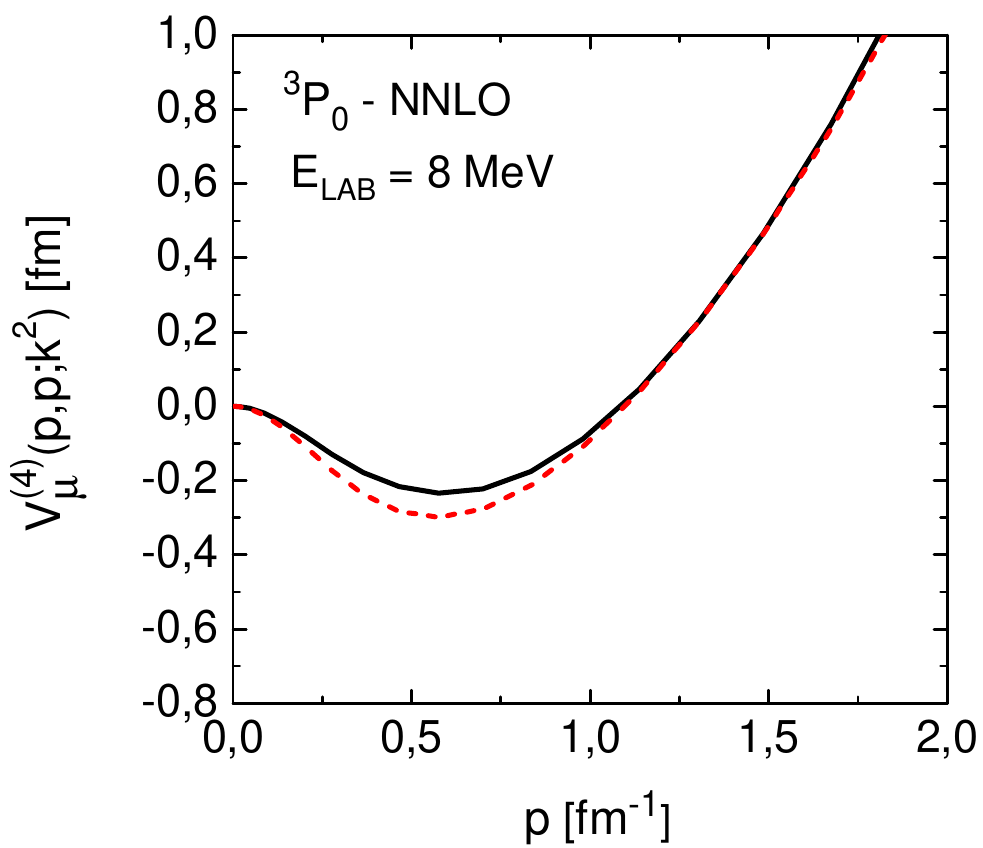}&\hspace{-0.3cm}
\includegraphics[scale=0.41]{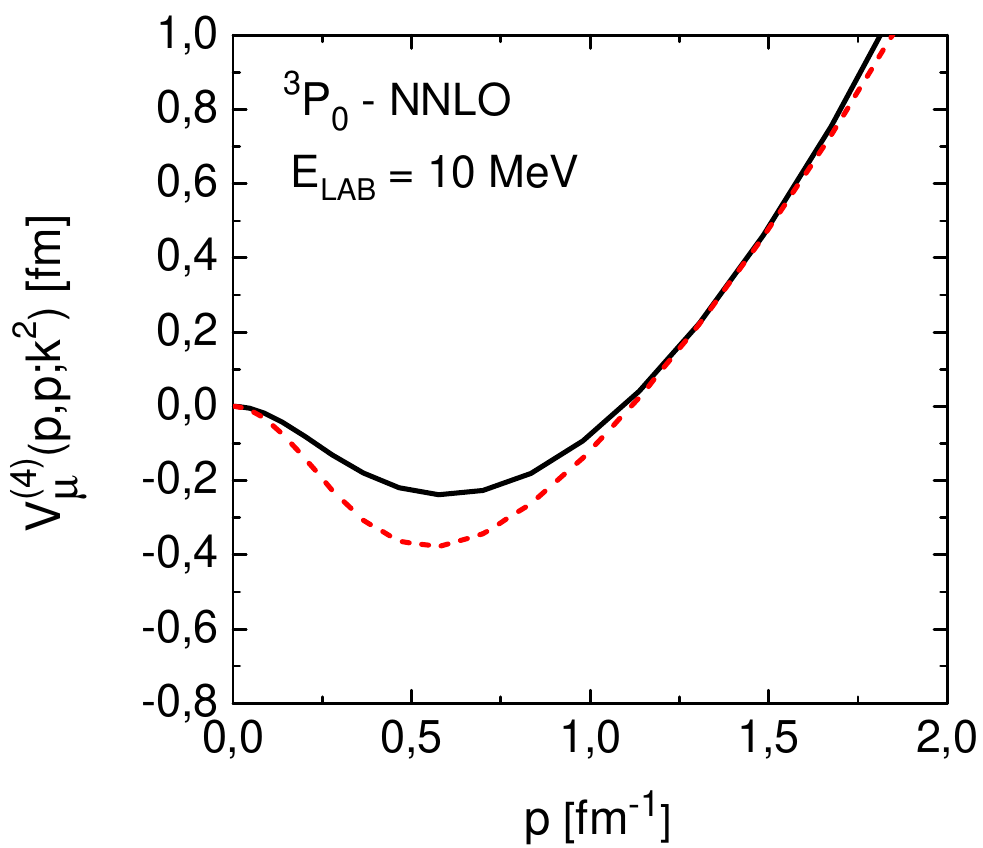}&\hspace{-0.3cm}
\includegraphics[scale=0.41]{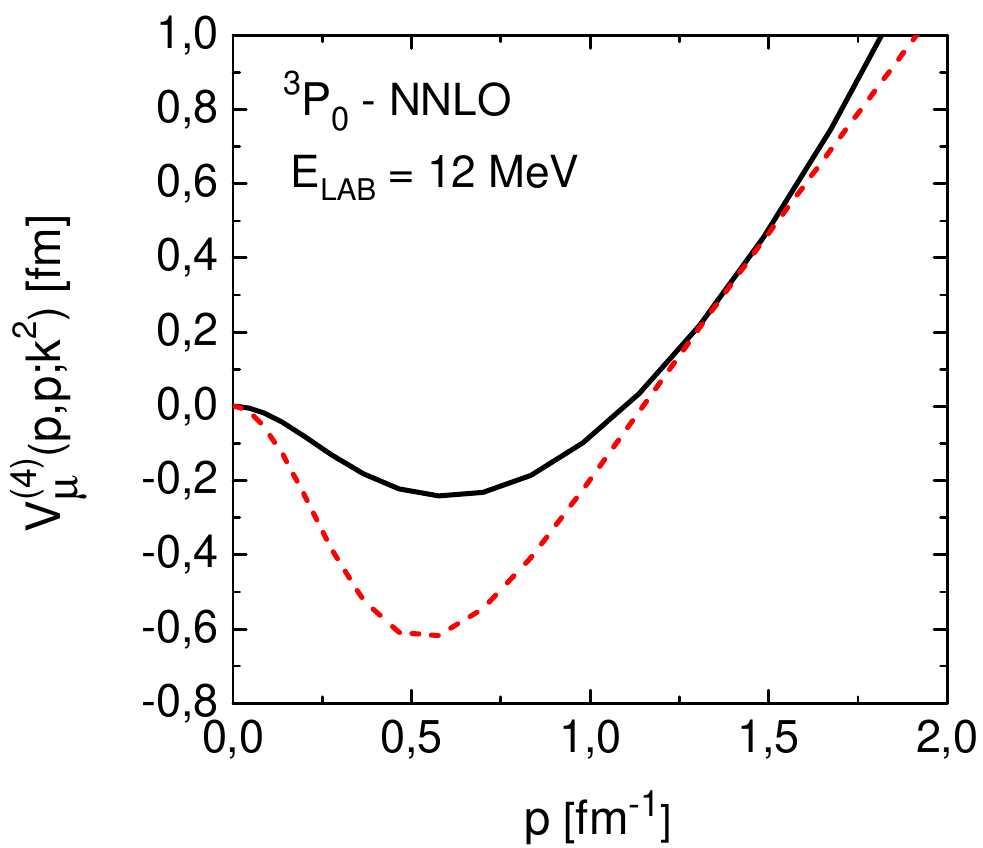}\\ \\
\includegraphics[scale=0.41]{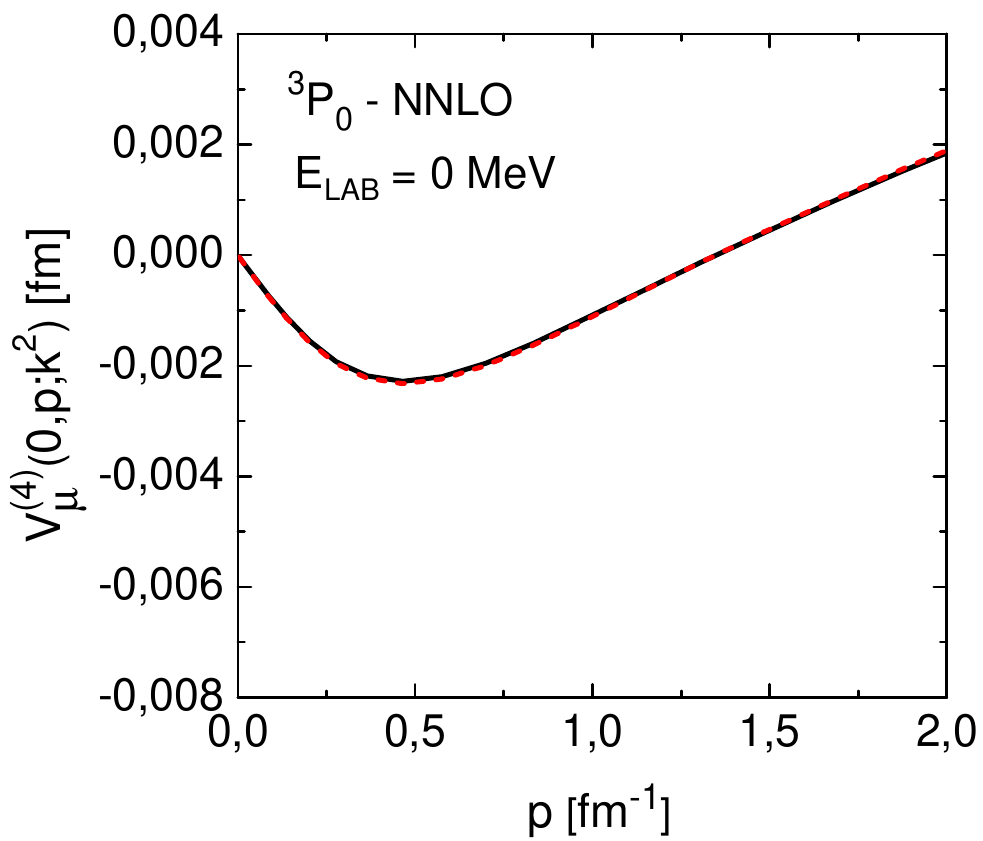}&\hspace{-0.3cm}
\includegraphics[scale=0.41]{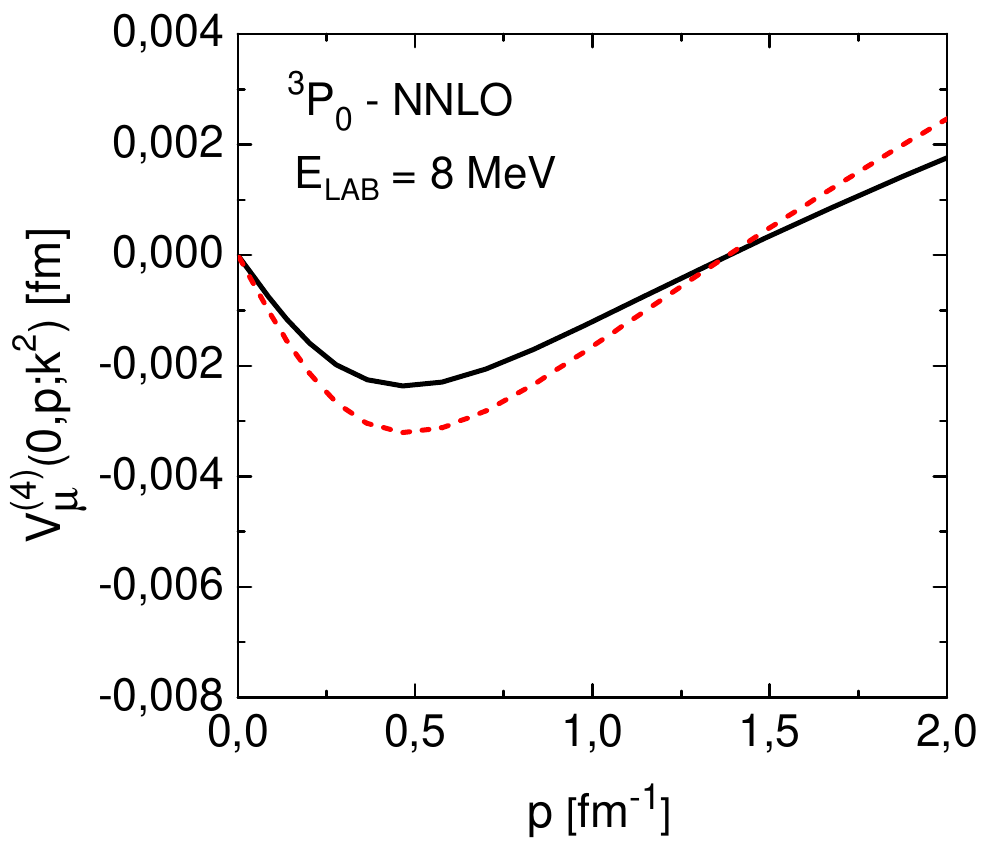}&\hspace{-0.3cm}
\includegraphics[scale=0.41]{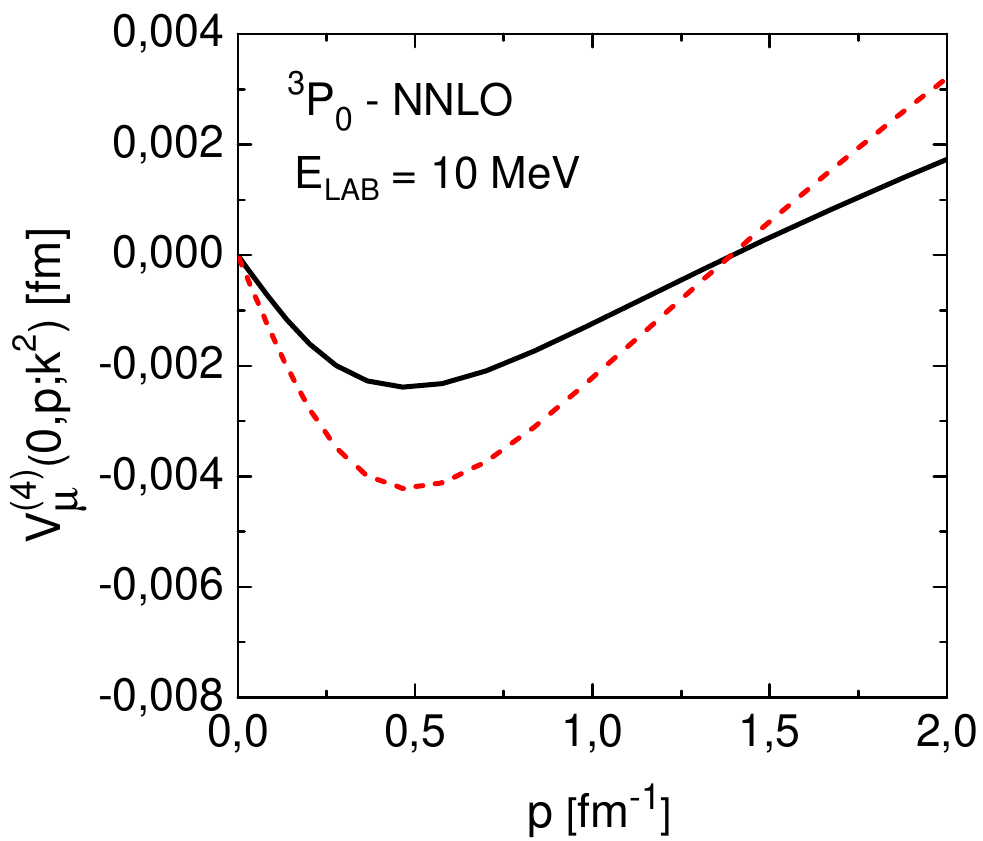}&\hspace{-0.3cm}
\includegraphics[scale=0.41]{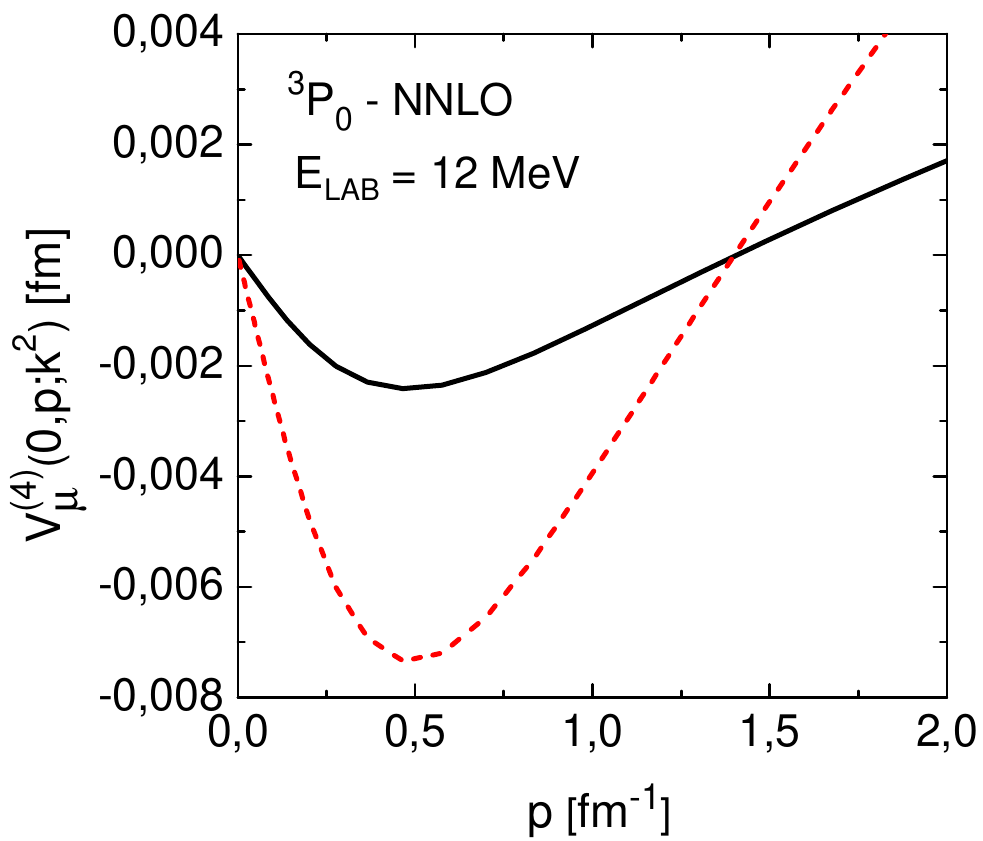}
\end{array}$
\caption{(Color on-line) Evolution through the NRCS equation of the $^3P_0$ channel driving term $V^{(4)}_{\mu}$ at $NNLO$ for several values of $E_{\rm LAB}$. Top panels: diagonal matrix elements; Bottom panels: off-diagonal matrix-elements.}
\label{fig1}
\end{figure}

We solve Eq. (\ref{CSEnPW}) numerically for $n=4$, obtaining an exact (non-perturbative) solution for the evolved $^3P_0$ channel driving term $V^{(4)}_{\mu}$. The relative momentum space is discretized on a grid of $200$ gaussian integration points, leading to a system of $200 \times 200$ non-linear first-order coupled differential equations which is solved using an adaptative fifth-order Runge-Kutta algorithm. In Fig.~\ref{fig1} we show the evolution of the diagonal matrix-elements (top panels) and the off-diagonal matrix-elements (bottom panels) of the $^3P_0$ channel driving term $V^{(4)}_{\mu}$ from a reference subtraction scale ${\bar \mu}=0.7~\rm{fm}^{-1}$ to $\mu=0.1~\rm{fm}^{-1}$ for several values of $E_{\rm LAB}$.

\begin{figure}[h]
\centerline{\includegraphics[scale=0.8]{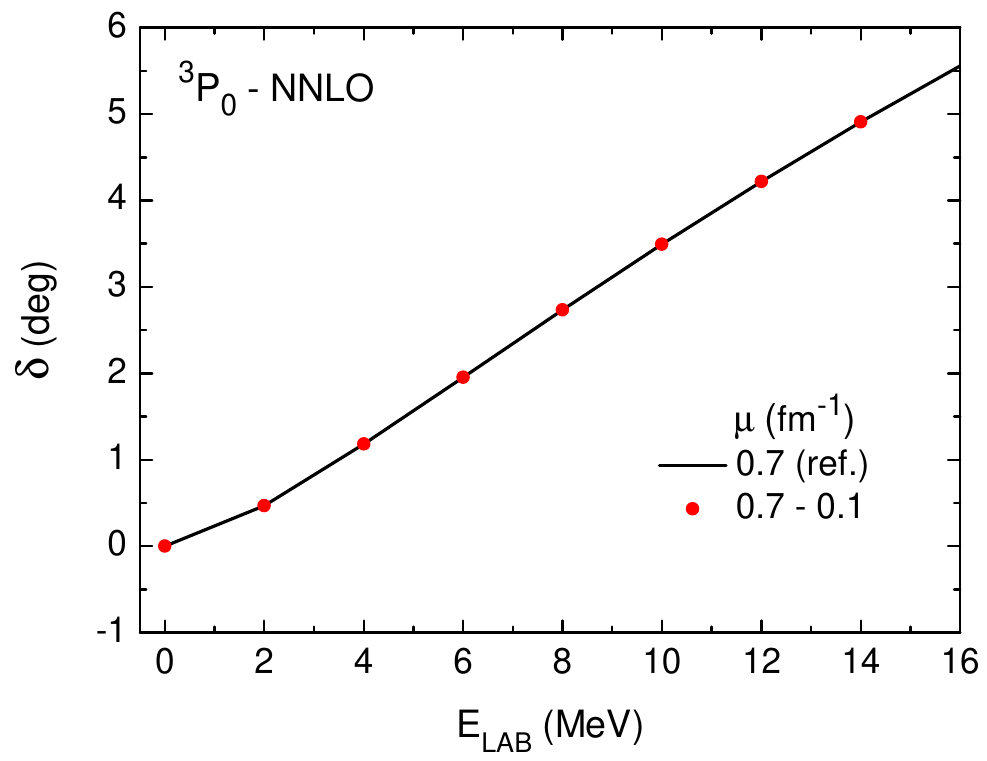}}
\caption{(Color on-line) Phase-shifts in the $^3P_0$ channel obtained with the driving term $V^{(4)}_{\mu}$ at $NNLO$ evolved through the NRCS equation. For comparison, we also show the phase-shifts obtained with $V^{(4)}_{\bar \mu}$ computed at the reference scale ${\bar \mu}$ (solid line).}
\label{fig2}
\end{figure}

As shown in Fig.~\ref{fig2}, the evolution of the driving term $V^{(4)}_{\mu}$ through the NRCS equation ensures that the phase-shifts calculated from the solution of the LS equation for the $4$-fold subtracted $K$-matrix remain invariant (except for relative differences smaller than $10^{-12}$ due to numerical errors).

\section{Summary and Conclusions}
In summary, we have demonstrated by explicit numerical calculations for the scattering of two-nucleons in the $^3P_0$ channel that the SKM procedure applied to the chiral $NN$ interactions up to $NNLO$ is renormalization group invariant under the change of the subtraction scale. Once the renormalized strength of the contact interaction is fixed at a reference scale to fit the generalized scattering length, the subtraction scale can be changed by evolving the driving term of the subtracted LS equation through a  non-relativistic Callan-Symanzik (NRCS) flow equation, such that the results for the calculated phase-shifts remain invariant. In this way, the sliding subtraction scale vanishes as a physical parameter. The relevant scale parameter left in the theory is the reference scale where the boundary condition of the NRCS equation is determined through the input of physical information.

\begin{theacknowledgments}
This work was supported by FAEPEX/PRP/UNICAMP and FAPESP. Computational power provided by FAPESP grants 2011/18211-2 and 2010/50646-6.
\end{theacknowledgments}

\bibliographystyle{aipproc}

\end{document}